\documentclass{article}

\usepackage{graphicx}
\usepackage[round]{natbib}
\usepackage{epsfig}

\title{
Predicting hurricane regional landfall rates:
comparing local and basin-wide track model approaches
}

\author{Tim Hall, GISS\footnote{\emph{Correspondence address}: Email: \texttt{tmh1@columbia.edu}}\\and\\
Stephen Jewson\\}

\begin{document}
\maketitle

\begin{abstract}
We compare two methods for making predictions of the climatological distribution of
the number of hurricanes making landfall along
short sections of the North American coastline.
The first method uses local data, and the second method uses a basin-wide track model.
Using cross-validation we show that the basin-wide track model gives better predictions for almost all parts of the coastline.
This is the first time such a comparison has been made, and is the first rigourous justification
for the use of basin-wide track models for predicting hurricane landfall rates and hurricane risk.
\end{abstract}

\section{Introduction}

There is considerable interest in trying to predict the number of
hurricanes that might make landfall on different parts of the
North American coastline in future years. Such predictions are
needed, for example, by insurance companies, to set insurance
rates, and local government, to set building codes. There are
various questions that need to be considered when making these
predictions, such as:

\begin{itemize}
    \item how is climate change affecting hurricane activity?
    \item which parts of the historical data are relevant to current and
    future hurricane activity?
    \item what methods give the most accurate predictions
    of future landfall activity?
\end{itemize}

Addressing these questions in detail is a challenging undertaking,
and the science of predicting hurricane landfall rates is still in
its infancy. Before trying to answer all these questions, it is
important to understand individual parts of the problem, and the pros
and cons of methodologies that one might use.
In this article we focus on an important basic question:
if we wish to estimate hurricane landfalling rates for a small section
of the coastline, is it better
to make that estimate using local data, or using a basin-wide
track model? Using local data is the simplest and most obvious
thing to do. For instance, to predict the number of hurricanes
making landfall in Texas, one might fit a distribution to the number
of hurricanes hitting Texas in the historical record.
However, this
method may not make the best use of available historical data,
since it ignores all the data for hurricanes that \emph{don't}
hit Texas, and this data might contain useful additional information that
could improve the estimates.
This reasoning has been one of the factors that has motivated the development of basin-wide
hurricane track simulation models, and these models are commonly
used by the various entities that need to understand hurricane risk.
However, in spite of there being
a number of publications describing how one might build such a
model (such as~\citet{chu98}, \citet{darling91}, \citet{drayton00}, \citet{emanuel05},
\citet{jamesm05}, \citet{rumpf}, \citet{velez}, \citet{vickery00}, our own work described in~\citet{hallj05f}, and others),
we are not aware of any serious attempt to
evaluate whether such models really work as a method for
estimating regional landfall rates. The purpose of this paper, therefore,
is to ask that question: can basin-wide track models give better
predictions of landfall rates than local estimates? To simplify
matters, we ask this question in the context of an assumption that
the climate was stationary for the period 1950-2003. This is
clearly not correct, as there have been well documented
interdecadal-timescale fluctuations in the numbers of hurricanes
during this period (see, for example,~\citet{goldenberg}).
However, making this assumption wouldn't seem
to benefit either the local or the basin-wide track model methods
we are comparing, and it makes the question of how to compare the two classes of model
more tractable.

What is the likely outcome of this comparison between the local and basin-wide track model methods?
On the one hand,
the basin-wide models use much more data to predict local landfall
rates: this is one of the main reasons for building such models, as mentioned above.
This might make them more accurate than the local methods.
On the other hand, no basin-wide model is likely to be perfect, and all are likely to have
biases relative to the real statistical behaviour of hurricanes.
If these biases are large, this could easily overwhelm the
benefits of using more data. Which of the local and basin-wide track model methods is better
therefore depends on a trade-off between these two effects.
We suspect that for very large
regions of coastline (for instance, for the entire North American
coastline) the use of local data is likely to be relatively more
successful. On the other hand, for small sections of coastline, especially those with
low hurricane landfall rates, the use of basin-wide track models
might be more successful, since the benefit of using
surrounding data is likely to be greater.

\emph{Prima facie} making a fair comparison between local methods and track
models would seem to be difficult: what is the standard against
which these two methods should be compared? This problem, is, however,
solved relatively easily using cross-validation. In other words, we split the
data, build the models on one part of the data, and compare the ability of the models
to predict the other part.
We use the best possible version of this approach, which is the leave-one-out jack-knife.
Since we are testing the ability to predict a
\emph{distribution}, and not just a single value, we need to use a probabilistic scoring system. We choose
what we think is the most sensible generic probabilistic score,
which is the out-of-sample expected log-likelihood. This score is the obvious
extension of Fisher's log-likelihood~\citep{fisher1912} to out-of-sample testing. It has been used
by a number of authors such as~\citet{dowe}, ~\citet{roulstons02} and~\citet{j94}.

\section{Methods}

Our goal is to compare local and track model methods for
estimating hurricane landfall rates. In particular, we will
compare the local methods described in~\citet{hallj06a} with the track model
of~\citet{hallj05f}.
Both types of model will be used for predicting the climatological distribution of the
number of hurricanes making landfall on each of 39 segments of the North
American coastline. These segments don't overlap, and together
they form an approximation for the whole North American
coast from the Yucatan peninsula in Mexico, to Canada.

\subsection{Local methods}
Our local method for predicting hurricane landfall rates comes from~\citet{hallj06a}:
it takes the historical data for the number of hurricanes crossing the coastline segment,
and fits a poisson distribution using Bayesian statistical
methods. The method
integrates over all possible poisson distributions,
which results in a form of the negative binomial
distribution. The main result in~\citet{hallj06a} is that this Bayesian method
works better than the obvious classical method of fitting just a single best-fit poisson distribution.
The improvement gained by using the Bayesian method is largest when the number of historical landfalls
in the segment is zero or very small.

\subsection{Track model methods}
The track model we use to predict landfall rates is described in a
series of six short articles (the sixth of which is~\citet{hallj05f}). It takes historical data
for historical hurricane genesis, tracks and lysis, and builds a statistical
model that can simulate an arbitrary number of future hurricanes.
The genesis model is poisson, and hence the distribution of numbers of hurricanes crossing coastal
segments in the model is also poisson.
As described in the papers cited above, we took particular care to build the model in a such a way that it
is not overfitted, and so it should stand a good chance of making
accurate predictions of landfalling rates.
We note, however, that the model is noticeably imperfect, and that landfall rates for certain
parts of the coastline are clearly wrong (see figure 7 in~\citet{hallj05f}).
We are working on eliminating such biases from the model, but for now we test the model
as is, warts and all.

\subsection{Comparison}

We compare these two models using leave-one-out cross-validation, using the out-of-sample expected
log-likelihood as the merit function, as follows.

\begin{itemize}

    \item We approximate the North American coastline using 39 straight line segments. We consider
    these segments one at a time.

    \item We loop over the 54 years of historical hurricane data from 1950 to 2003
    (this data is all taken from the 2004 version of HURDAT~\citep{hurdat}).

    \item We miss out each year in turn

    \item Using the remaining years, we fit both models.
    For the local model, fitting consists of simply estimating the model
    parameters using the historical landfalling data for that segment. For the track model, fitting
    consists of estimating the smoothing length-scales in each part of the model using the entire
    historical hurricane database.
    In fact, the smoothing lengthscales
    turn out not to change when we miss out single years of data, since they are only calculated at
    a fairly low resolution, and all 54 fitted track models have the same smoothing lengthscales.

    \item For each model, we then make a prediction of the number of hurricanes making
    landfall in the year missed out, for the segment of interest.
    For the local models, the prediction is simply the fitted
    distribution. For the track model, the prediction is based on counting storms in a 500 year
    simulation from the model, and fitting a poisson distribution
    (using the same Bayesian method as used for the local model). The simulations do not
    use any information from the year being predicted.

    \item We then compare the predictions for each year with the actual number of hurricanes in each year.
    We do this by evaluating the predicted distribution function at the observed number of storms,
    which gives a single probability density value.

    \item We take the log of this probability density and create an average of the 54 log scores to give
    an overall score for each model for each segment.

    \item We repeat this whole exercise for each of our 39 segments.

\end{itemize}

This gives us a single score for each model for each segment. We can then compare these scores between the two models.

\section{Results}

\begin{figure}[!hb]
  \begin{center}
    \scalebox{0.7}{\includegraphics{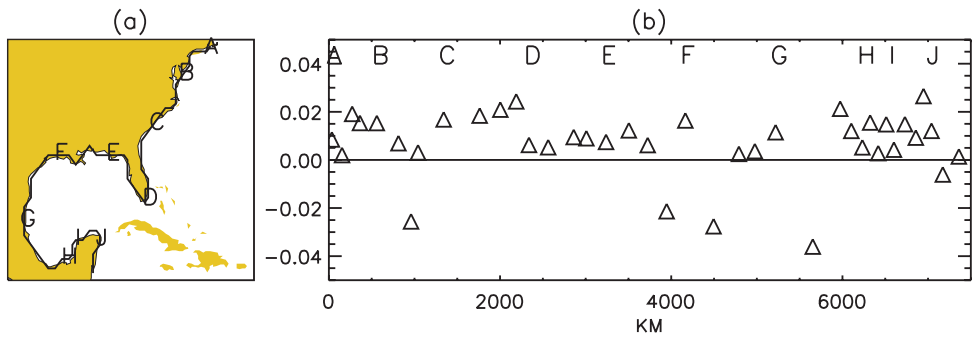}}
  \end{center}
    \caption{
The left hand panel shows the eastern coastline of North America, with letters as reference points.
The right panel shows the difference in predictive performance of the two models we have tested, for
hurricane landfall rates along this coastline. Positive values indicate that the basin-wide track model
works better, while negative values indicate that the local model works better. We see that the basin-wide
track model wins for 34 of the 39 segments.
}
     \label{f01}
\end{figure}

The results of this score comparison are shown in figure~\ref{f01}.
The left panel shows the definitions of the coastal segments used in the comparison.
The right panel shows the expected log-likelihood score for the basin-wide model
minus the score for the local model. Positive values indicate that the basin-wide model
is giving more accurate predictions than the local model.

We see that the track model wins this comparison for 34 of the 39 coastal segments.
One can ask whether this could happen by chance, if the two models were in fact equally good. Using a statistical
test based on the binomial distribution, it turns out that this would be \emph{extremely} unlikely to happen by chance.
As a result we can conclude that the track model is genuinely better (using this score) than the
local method, on average over all the segments, but not for every segment individually.

It is interesting to compare these results with the results shown in figure 7 in~\citet{hallj05f}, which shows
an \emph{in-sample} comparison between landfall rates from the track model and landfall rates from the local model.
For four of the five gates where the track model loses in the current study,
we can see that problems with the track model
already show up in these in-sample comparisons.
For instance, in~\citet{hallj05f} the track model gives lower landfall rates than the local method would predict between B and C, between E and F, and
between F and G, and the current study shows that this is probably because the track model is wrong in those regions.
The track model gives higher landfall rates than the local method between G and H, and again, the current
study shows that that is probably because the track  model is wrong.
On the other hand, the track model gives lower rates than the local method between H and I, but the current
study suggests that the track model is actually more accurate for this part of the coastline:
one might say that this implies that the region between H
and I has been unlucky during this 54 year period, and experienced more hurricanes than it would do on average.
Similarly at point E the track model gives higher rates than
the local method, but the current study suggests again that the track model is more accurate for this location.
In this case one might say that point E has been lucky over the last 54 years, and experienced fewer
hurricanes than it would be expected to on average.

\section{Discussion}

Regional landfalling rates for hurricanes are often estimated either using local methods, that count the number
of hurricanes that have hit a region in the historical record, or using track model methods,
that count the number of hurricanes that hit that region in statistical simulations.
However, in spite of considerable effort to develop a number of different track models
(see citations to 9 different models in the introduction)
there has never been any rigourous
attempt to compare the two types of method (or indeed, to compare the track models with each other).
One might hope that the track models will do better since they use more data than the local methods, but
whether they \emph{really} do perform better or not depends on the size of the (inevitable) errors in the track
models, and can't be guessed in advance: \emph{it is purely a matter for statistical testing}.
To be an honest comparison such testing can only be performed using cross-validation, since in-sample testing
favours over-fitted models with many parameters. This point is particularly important because
track-models are all based on a very large number of fitted parameters.

We perform such a test for the first time, by comparing local methods based on Bayesian
statistics from~\citet{hallj06a} with the track model of~\citet{hallj05f}.
We find that the track model gives a better estimate of the climatological distribution of the number of hurricanes
making landfall for 34 out of 39 coastal gates.
We consider this result to be a major milestone in the development of methods for
predicting rates of landfalling hurricanes:
for the very first time we have shown that track models can give more accurate predictions than local methods.
This is, on its own, a justification for the whole enterprise of trying to build track models.
We do emphasize, however, that we have not shown that \emph{all} track models necessarily work better
than the local method, but only the particular track model that we have tested.

This result opens up a number of directions for future study, such as:
\begin{itemize}
   \item would the same track model also work better than local methods for larger coastline segments? Up to what limit?
   \item would the same track model also work better for the prediction of intense storms?
   \item could the performance of the track model be further improved, to the extent that it would
   beat the local method for all 39 coastal segments?
   \item could other track models, such as the 8 other models cited in the introduction, also beat the local methods?
   \item of all the published track models, which is the best in terms of landfall rate predictions?
\end{itemize}

We plan to look at all these questions in due course.

\bibliography{arxiv}

\begin{thebibliography}{16}
\providecommand{\natexlab}[1]{#1}
\providecommand{\url}[1]{\texttt{#1}}
\expandafter\ifx\csname urlstyle\endcsname\relax
  \providecommand{\doi}[1]{doi: #1}\else
  \providecommand{\doi}{doi: \begingroup \urlstyle{rm}\Url}\fi

\bibitem[Chu and Wang(1998)]{chu98}
P~Chu and J~Wang.
\newblock Modelling return periods of tropical cyclone intensities in the
  vicinity of {H}awaii.
\newblock \emph{Journal of Applied Meteorology}, 39:\penalty0 951--960, 1998.

\bibitem[Darling(1991)]{darling91}
R~Darling.
\newblock Estimating probabilities of hurricane wind speeds using a large-scale
  empirical model.
\newblock \emph{Journal of Climate}, 4:\penalty0 1035--1046, 1991.

\bibitem[Dowe et~al.(1996)Dowe, Farr, Hurst, and Lentin]{dowe}
D~Dowe, G~Farr, A~Hurst, and K~Lentin.
\newblock Information-theoretic football tipping.
\newblock In N~deMestre, editor, \emph{3rd conference on mathematics and
  computers in sport}, pages 233--241, 1996.

\bibitem[Drayton(2000)]{drayton00}
M~Drayton.
\newblock A stochastic basin-wide model of {A}tlantic hurricanes, 2000.
\newblock 24th Conference on Hurricanes and Tropical Meteorology.\\
  \texttt{http://ams.confex.com/ams/last2000/24Hurricanes/abstracts/12797.htm}.

\bibitem[Emanuel et~al.(2005)Emanuel, S, E, and C]{emanuel05}
K~Emanuel, Ravela S, Vivant E, and Risi C.
\newblock A combined statistical-deterministic approach of hurricane risk
  assessment.
\newblock Unpublished manuscript, 2005.

\bibitem[Fisher(1912)]{fisher1912}
R~Fisher.
\newblock On an absolute criterion for fitting frequency curves.
\newblock \emph{Messenger of Mathematics}, 41:\penalty0 155--160, 1912.

\bibitem[Goldenberg et~al.(2001)Goldenberg, Landsea, Mestas-Nunez, and
  Gray]{goldenberg}
S~Goldenberg, C~Landsea, A~Mestas-Nunez, and W~Gray.
\newblock {The recent increase in Atlantic hurricane activity: causes and
  implications}.
\newblock \emph{Science}, 293:\penalty0 474--479, 2001.

\bibitem[Hall and Jewson(2005)]{hallj05f}
T~Hall and S~Jewson.
\newblock Statistical modelling of tropical cyclone tracks part 6: non-normal
  innovations.
\newblock \emph{arXiv:physics/0512135}, 2005.

\bibitem[Hall and Jewson(2006)]{hallj06a}
T~Hall and S~Jewson.
\newblock {Comparing classical and Bayesian methods for predicting hurricane
  landfall rates}.
\newblock \emph{arXiv:physics/0611006}, 2006.

\bibitem[James and Mason(2005)]{jamesm05}
M~K James and L~B Mason.
\newblock {Synthetic tropical cyclone database}.
\newblock \emph{{Journal of Waterways, Coastal and Ocean Engineering}},
  131:\penalty0 181--192, 2005.

\bibitem[Jarvinen et~al.(1984)Jarvinen, Neumann, and Davis]{hurdat}
B~Jarvinen, C~Neumann, and M~Davis.
\newblock {A tropical cyclone data tape for the North Atlantic Basin,
  1886-1983: Contents, limitations, and uses}.
\newblock Technical report, {NOAA Technical Memorandum NWS NHC 22}, 1984.

\bibitem[Jewson and Penzer(2006)]{j94}
S~Jewson and J~Penzer.
\newblock Weather derivative pricing and the normal distribution: fitting the
  variance to maximise expected predictive log-likelihood.
\newblock \emph{http://ssrn.com/abstract=911569}, 2006.

\bibitem[Roulston and Smith(2002)]{roulstons02}
M~Roulston and L~Smith.
\newblock Evaluating probabilistic forecasts using information theory.
\newblock \emph{Mon. Wea. Rev.}, 130:\penalty0 1653--1660, 2002.

\bibitem[Rumpf et~al.(2006)Rumpf, Rauch, Schmidt, and Weindl]{rumpf}
J~Rumpf, E~Rauch, V~Schmidt, and H~Weindl.
\newblock {Stochastic modelling of tropical cyclone track data}, 2006.
\newblock 27th conference of hurricanes and tropical meteorology.

\bibitem[Velez et~al.(2005)Velez, Lall, Rajagopalan, and Kushnir]{velez}
J~Velez, U~Lall, B~Rajagopalan, and Y~Kushnir.
\newblock {A Markov and track segment model for simulating hurricane risk with
  Atlantic Ocean applications}, 2005.
\newblock Poster at \emph{Tropical Cyclones and and Climate Workshop, IRI,
  Columbia University}.

\bibitem[Vickery et~al.(2000)Vickery, Skerlj, and Twisdale]{vickery00}
P~Vickery, P~Skerlj, and L~Twisdale.
\newblock Simulation of hurricane risk in the {US} using an empirical track
  model.
\newblock \emph{Journal of Structural Engineering}, 126:\penalty0 1222--1237,
  2000.

\end{thebibliography}

\newpage

\end{document}